\documentclass[prx,noshowpacs,print,twocolumn]{revtex4}
\usepackage{pifont}
\usepackage{amssymb}
\usepackage{amsmath}
\usepackage{graphicx}
\usepackage{dcolumn}
\usepackage{bm}
\usepackage{txfonts}
\usepackage{appendix}
\usepackage[dvips]{color}
\usepackage{bm}
\usepackage{multirow}
\begin{document}

\title{Crystallographic Characterization of Black Phosphorene and Its Application in Nanostructures}
\author{Yi Ren, Pu Liu, Benliang Zhou, Xiaoying Zhou,\footnote{xiaoyingzhou@hunnu.edu.cn}
and Guanghui Zhou\footnote{ghzhou@hunnu.edu.cn}}
\affiliation{Department of Physics and Key Laboratory for
Low-Dimensional Quantum Structures and Manipulation (Ministry of
Education), and Synergetic Innovation Center for Quantum Effects and
Applications of Hunan, Hunan Normal University, Changsha 410081,
China}

\begin{abstract}
The central question in the field of two-dimensional (2D) materials
is how a material behaves when it is patterned at nanometer scale
with different edge geometries. Due to the anisotropy inherent in
the puckered structure, black phosphorene nanostructures may have
more varieties of edge geometries. Here, we present a comprehensive
2D planar crystallographic characterization of phosphorene uniformly
by a chiral vector (angle), from which a new type of edge atomic
configuration, the slope edge geometry is discovered. The
phosphorene nanoribbons (PNRs) with slope edges, like previously
noticed zigzag and skewed-armchair PNRs, also own interesting
twofold-degenerate edge states. These three marginal directions,
together with the skewed-zigzag and armchair directions without edge
states, divide a phosphorene into four regions among which the
electronic property is different from each other. Further, for a PNR
cutting along any possible direction, by taking into account the
$z$-direction in structure, whether or not it owns edge states
depends on the existence of periodic zigzag-like morphology along
the edges. For application, moreover, we propose a PNR-based
z-shaped homogenous junction with scale $\sim$100 nm, where the
central scattering region between two PNR electrodes is a
phosphorene quantum dot (PQD) of PNR segment with various edge
morphologies. Interestingly, the calculated conductance by Kwant
code based on tight-binding combing with scattering-matrix for the
junction relies sensitively on the central PQD edge states. In
specification, for the junctions of PQD with edge states the
conductance exhibit a terrace with resonant oscillations near Fermi
energy, otherwise the electron transport is blocked due to the
absence of edge states. Remarkably, the number of oscillating peaks
exactly matches to the number of sawtooth along an edge of PQD,
since the discrete energy levels of the zigzag-like edge provide the
transporting channels for electrons. The results here may be
extended to the group-VA 2D materials for observing the electronic
property of nanostructures in experiments, and provide a reference
for the preparation of better nano-devices.
\end{abstract}

\maketitle

\section{Introduction}
In this decade, two-dimensional (2D) materials have emerged as one
of the most exciting class of materials since the arising of
graphene \cite{Geim}. Recently, semiconducting black phosphorous (or
phosphorene), a mono- or few-layer of phosphorus atoms forming 2D
puckered sheets, has joined the class of 2D materials \cite{Li,
LiuH, XiaF, Steven, Buscema, Ling}. The synthesized phosphorene has
a noticeable band gap and appreciable anisotropy in its physical
properties \cite{Ling, Rishabh, Chen}. Generally, a 2D material may
display the properties differing greatly from its bulk counterpart
due to the effects of quantum confinement and large
surface-to-volume ratio. Further, tailoring a 2D material along its
different crystallographic directions can obtain one-dimensional
(1D) nanoribbons with the width less than 100 nm \cite{Tao, LiY,
Mitchell, Carvalho, Marko}. These nanoribbons show not only the
confinement effect but also peculiar edge effect, both of which play
an important role in electronic, optical, magnetic and transport
properties \cite{SGLouie, Tran, Ajanta, YangG}. Hence the edge
states are particularly important in a 2D material nanoribbon.

Black phosphorus, with an anisotropic structure, is a van der
Waals-bonded layered material where each layer forms a puckered
surface due to \emph{sp}$^3$ hybridization and possesses a direct
band gap of 0.3 eV \cite{HanCQ}. This direct gap increases up to
$\sim$2 eV as its thickness decreases from bulk to few layers
\cite{Li1}. In addition, the field-effect transistor (FET) \cite{Li,
Steven, Buscema} based on phosphorene is found to have an on/off
ratio of $10^{3}$ and a high hole carrier mobility up to 800
cm$^{2}$/V$\cdot$s \cite{Sherman}. Importantly, there have been a
few effective ways to solve the instability problem for phosphorene
in the environment \cite{Niu}. These good properties of phosphorene
have attracted a broad interests \cite{Zhou, Zhou1, Li2, Quhe,
Gautam}. Further, PNRs by tailoring a phosphorene sheet in the
conventional zigzag and armchair directions have been prepared
experimentally \cite{Paul, Mitchell}, respectively. The zigzag PNRs
(ZPNRs) with significant edge states and armchair PNRs (APNRs) with
direct band-gap have been extensively studied \cite{Carvalho, Tran,
Ezawa, Ajanta}. And some properties of PNRs have been verified
\cite{YangG, Esmaeil, Sousa, SK, Ren, Sisakht, ZhouBL, Han}, such as
the effect of edge states on conductance and thermalpower \cite{Ma},
the influence of vacancy defects on electron structure and transport
\cite{Zahra, Lill}, and the electric tunable edge-pseudospin valve
\cite{Soleimanikahnoj}.

On the other hand, the nanostructures with various possible edge
atomic configurations have become a hot topic since the rise of
2D materials. A central question in the field of graphene-related
research is how it behaves as patterned at the nano-scale with
different edge morphological geometry \cite{Tao, Akhmerov}. If we
cut a graphene sheet along crystal directions by increasing the
chiral angle from 0$^{\circ}$ to 30$^{\circ}$, for example, the edge
geometry switches from zigzag to armchair, between which we can
obtain all possible chiral ribbons with different chiral angles
\cite{Tao, Akhmerov}. Recently, the effects of chirality and edge
geometry on the anisotropic thermal transport property have also
been studied for MoS$_2$ nanoribbons \cite{LiuT}. For PNRs, in
addition to the conventional zigzag and armchair edges, other edge
geometries have been predicted. For example, the PNRs with stable
edges of beard-zigzag \cite{Ezawa}, skewed-zigzag (SZ) and
skewed-armchair (SA) \cite{Marko, Ashwin, Liu} have also been
reported. And some of them have already been realized experimentally
\cite{Liang}. The electronic property, essentially, is significantly
different from each other among PNRs with different chiralities.
This difference results in abundant electrical nature, such as metal
with edge states and semiconductor with direct band gap.
Furthermore, the difference in physical properties has potential
applications. For instance, the control of external electric and
magnetic fields on the electronic structure \cite{Vladimir} and the
coupling and manipulation of the edge states in multilayer PNRs
\cite{Lv}. However, the existed studies have only involved PNRs with
above mentioned few particular edge geometries. An uniform
verification of the edge morphology for highly anisotropic
phosphorene, as those for isotropic graphene \cite{Tao, Akhmerov},
is also important in the phosphorene-based nanotechnology.

\begin{figure}[t]
\centering
\includegraphics[bb=5 2 573 396,width=8.5cm]{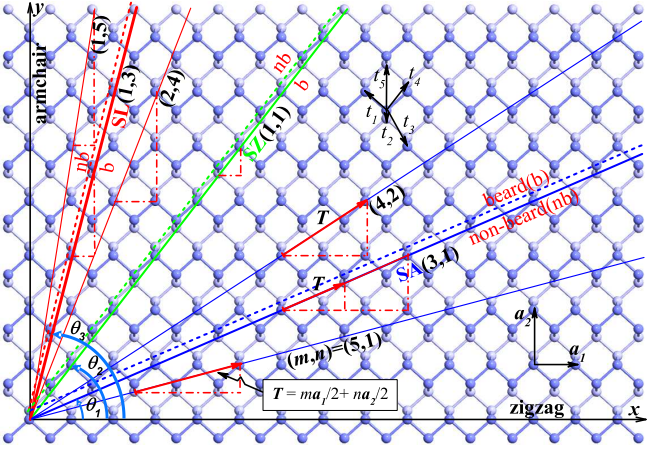}
\caption{The illustration of crystallographic
characterization on a monolayer phosphorene, where the heavy (light)
gray spheres donate the phosphorus atoms in the upper (lower)
subplane, and the $x$- and $y$-coordinates are respectively along
the zigzag and armchair crystallographic directions. The (blue),
(green) and (red) thick solid lines along the SA-, SZ- and
SL-directions with chiral angles $\theta_1$, $\theta_2$ and
$\theta_3$, respectively; $\bm{a}_1$ and $\bm{a}_2$ denote the
primitive vectors; $t_1$-$t_5$ are the five hoping parameters. The
chiral vector $\bm{T}$ is illustrated by the (red) thick solid arrow
along each 2D crystallographic direction, where the arrows along
(4,2) and (3,1) show the examples of the outmost atoms are located
at the same sublayer and alternatively at two sublayers,
respectively.}
\end{figure}

In this work, we first present theoretically the 2D planar
crystallographic characterization on an anisotropic phosphorene. We
define a chiral number (vector) or chiral (azimuthal) angle to
uniformly describe the planar crystallographic directions. As the
chiral angle increasing from 0$\textordmasculine$ to
90$\textordmasculine$ shown in Fig. 1, the boundary morphology
switches from zigzag to armchair and sweep over all possible edge
geometries during the process. We find a phosphorene can be divided
into four fan-shaped regions determined by chiral angles
0$\textordmasculine$, $\theta_1(23.74\textordmasculine)$,
$\theta_2(52.84\textordmasculine)$,
$\theta_3(75.82\textordmasculine)$ and 90$\textordmasculine$, among
which the electronic characteristics is different. Therefore, by
cutting a phosphorene along these five regional boundary directions,
one can obtain previously studied ZPNR, SAPNR, SZPNR, and APNR
\cite{Marko, Ashwin, Liu}. Importantly, the PNR with sloped (SL)
edge geometry along $\theta_3$ (named as SLPNR) between SZ and
armchair direction has not been reported yet. Essentially, the SL
direction is similar to SA with a periodic zigzag-like edge
morphology supporting edge states. Further, by taking into account
the effect of the $z$-direction in structure, we then demonstrate
the variation of electronic property for PNRs with different edge
atomic geometries by the Kwant code based on tight-binding (TB)
calculations \cite{Groth}, in which a particular attention is paid
to the new SLPNR intrinsically with twofold-degenerate edge states.
In addition, the origin of beard and non-beard edges along each
regional boundary direction is discussed and related to the
existence of edge states. The result here may be extend to the
cousins of phosphorene, such as 2D group-VA arsenene, antimonene and
bismuthene \cite{ZhangSL}.

Second, based on the 2D crystallographic characterization, we
propose a two-terminal z-shaped homogenous junction based on PNR and
calculate its transport property using the Kwant code based on TB
\cite{Groth} combined with scattering matrix approach from
wavefunction-matching \cite{Yongjin,Zwierzycki,Tatiane} for treating
large enough systems comparable with experimental samples. The two
electrodes of the junction are semi-infinitive ZPNRs with the same
width $\sim$100 nm, where the central scattering region of a segment
of PNR with edge tilting angles from 0$\textordmasculine$ to
90$\textordmasculine$ is sandwiched between them. In consequence the
two hypotenuse edges of the central phosphorene quantum dot (PQD)
sweep over all possible geometrical morphologies. Interestingly, the
conductance spectrum sensitively depends on the tilting angle of
$z$-shaped junction. For the junctions of PQD with edge states the
transport behavior exhibit a conductance terrace with resonant
oscillation in the low energy regime, otherwise the electron
transport is blocked due to the absence of edge states. Remarkably,
the number of conductance oscillating peaks is exactly equal to the
number of outermost phosphorus atoms in zigzag-like edges of the
center PQD, since the discrete energy levels of the zigzag-like
atomic configurations provide the transport channels for electrons.
These findings may provide a reference for experimental preparation
of better nano-devices.

This paper is organized as follows. Sec. II we first present the 2D
crystallographic characterization of phosphorene, in which the
crystallographic direction on the plane is uniformly described by a
chiral vector or angle. By this vector (angle) the crystallographic
directions with possible boundary edge geometries are identified,
particularly including the newly recovered slope edge. Further, the
calculation method by using the Kwant code based on tight-binding
(TB) combing with scattering matrix approach is described. In Sec.
III, firstly a few examples of typical PNRs with stable structure
are analogized, and then the PNR-based z-shaped junction with
different tilting degrees, in which the central PQD with various
edge morphologies, is constructed, and the transport property is
studied in details. And the main results are presented numerically
with the microscopic explantation and discussion. Finally, we
summarize our results in Sec. IV.

\section{Model Description and Method}
\subsection{2D crystallographic characterization}
Phosphorene is a typical orthorhombic-structured 2D material with
two principal crystallographic directions, the zigzag and armchair
directions, as shown in Fig. 1, denoted as the $x$-axis and
$y$-axis, respectively. It is known that the period of minimum
rotation angle for graphene is 30$^{\circ}$ from zigzag to armchair
direction \cite{Tao, Akhmerov}. Due to the highly anisotropic
structure of phosphorene, however, it needs to anticlockwise rotate
90$^{\circ}$ from the zigzag to armchair direction. In consequence,
there may exist more possible directions along which form PNRs by
cutting a phosphorene in addition to along the conventional zigzag
and armchair directions, as well as the previously verified SA and
SZ directions \cite{Marko, Liu}. Therefore, an unified
identification for the crystallographic direction on 2D phosphorene
is still in great demand.

\begin{figure}[b]
\centering
\includegraphics[bb=86 45 731 532, width=8.5cm]{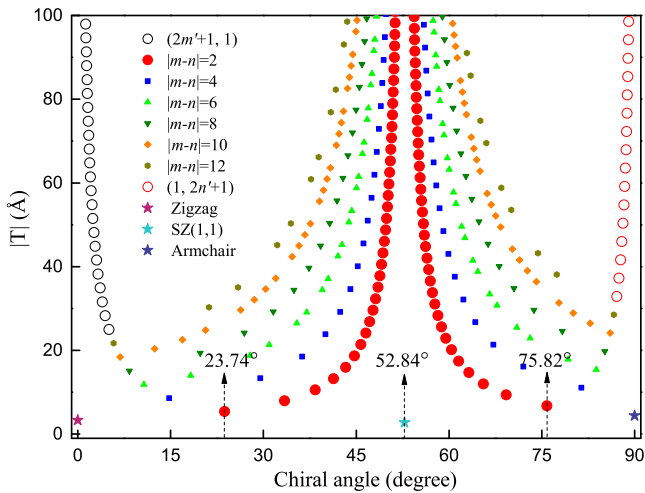}
\caption{The length of unitary chiral vector $\bm{T}$ as a function
of the chiral angle, where the lengths indicated by the (red)
spheres plus the (black) and (red) circles form a distributional
pattern with two separated U-shape in which all the other dots are
inside the valleys except for the three colored stars corresponding
to chiral angles 0, 52.84 and 90$\textordmasculine$, respectively.}
\end{figure}

The low symmetry of phosphorene leads to higher structural
complexity compared to other 2D materials \cite{Tao, Akhmerov,
LiuT}, such as graphene, silicene, $h$-BN, and MoS$_2$ which all
have a regular hexagonal symmetry in 2D. In the previous works, the
symmetry of phosphorene has been studied in terms of the point group
theory \cite{LiPK}. In order to simply characterize the
crystallographic directions for phosphorene by using the inversion
symmetry to reduce the degree of freedom \cite{Ezawa,LiPK}, here we
firstly ignore the $z$-direction in structure and add its effect on
nanostructures in the next step. This simplification means that
there are only two atoms in a primitive cell instead of the standard
four atoms per primitive cell. Therefore, the 2D planar
crystallographic directions of a phosphorene can be characterized by
an orientational (chiral) index $(m,n)$, or equivalently, a chiral
angle $\theta$. Further, $(m,n)$ or $\theta$ also determines a
minimum unit vector along the corresponding crystallographic
direction. Therefore, we can define a chiral vector to characterize
the crystallographic directions, which is given by
\begin{equation}
\bm{T}=m\bm{a}_{1}/2+n\bm{a}_{2}/2.
\end{equation}
Here $\bm{a}_1$ and $\bm{a}_2$ are the primitive vectors of
phosphorene, with amplitudes $a_1$=3.32 $\AA$ and $a_2$=4.38 $\AA$
along the $x$- and $y$-directions \cite{Rudenko}, respectively. The
module of $\bm{T}$ is the hypotenuse length of the (red) dot-dashed
line right-angled triangle for each direction shown in Fig. 1. Hence
\emph{m} and \emph{n} are integers for the projection of $\bm{T}$
onto $x$- and $y$-directions in terms of $a_1$/2 and $a_2$/2,
respectively. In order to ensure both $m$ and $n$ are integers, the
two sides of right-angled triangle in Fig. 1 are taken as half of
the primitive vectors. Thus, we can obtain a series of possible
crystallographic directions of phosphorene only when
\emph{m}+\emph{n}=even because the \emph{m}+\emph{n}=odd case is
unphysical according to our definition of $\bm{T}$. To satisfy the
condition for the chiral index, both \emph{m} and \emph{n} must
simultaneously be even or odd. As a result their difference is also
an even integer. Particularly, $|m-n|$=0 corresponds to chiral index
$(m,n)$=(1,1) for the SZ direction. In addition, when the difference
is a nonzero even number, the corresponding chiral index is (3,1),
(1, 3) and (5, 1), and so on, respectively (also see Fig. 1).

Meanwhile, we can accordingly introduce a chiral angle as
\begin{equation}
\theta=\arctan(n a_2/m a_1),
\end{equation}
with which connects the chiral index $(m,n)$. Therefore, a
crystallographic direction on the phosphorene plane can be
characterized by either a chiral vector $\bm{T}$ or index $(m,n)$,
or angle $\theta$. Despite all this, however, the structural
characterization is still much different from that for graphene with
a regular hexagonal symmetry \cite{Akhmerov}.

\begin{table}
\caption{The 2D crystallographic directions of phosphorene with
chiral indexes and angles on the two U-lines in Fig. 2.}
\begin{center}
\renewcommand{\arraystretch}{1.3}
\fontsize{7.5}{8}\selectfont
\begin{tabular}{cccccccccc}

\hline $0\textordmasculine-\theta_1$ &  & $\theta_1-\theta_2$ &  & $\theta_2-\theta_3$ &  & $\theta_3-90\textordmasculine$ &&\\
\hline zigzag-SA &  & SA-SZ &  & SZ-SL &  & SL-armchair &&\\
\hline $(m,n)$ $\theta$ & & $(m,n)$ $\theta$ & & $(m,n)$ $\theta$ &
& $(m,n)$ $\theta$ & &\\ \hline
(2,0) 0$\textordmasculine$ & & (3,1) 23.74$\textordmasculine$ & & (1,1) 52.84$\textordmasculine$ & & (1,3) 75.82$\textordmasculine$ &  &\\
(2$m'$+1,1)$\rightarrow$0$\textordmasculine$ & & (4,2)
33.41$\textordmasculine$ & &
($m'$,$n'$+2)$\rightarrow$52.84$\textordmasculine$ & & (1,5)
81.38$\textordmasculine$ &  &\\
$\vdots$ & & $\vdots$ & & $\vdots$ & & $\vdots$ &  &\\
(5,1) 14.78$\textordmasculine$ & & ($m'$+2,$n'$)$\rightarrow$52.84$\textordmasculine$ & & (2,4) 69.24$\textordmasculine$ & & (1,2$n'$+1)$\rightarrow$90$
\textordmasculine$ & &\\
(3,1) 23.74$\textordmasculine$ & & (1,1) 52.84$\textordmasculine$ & & (1,3) 75.82$\textordmasculine$ & & (0,2) 90$\textordmasculine$ & &  \\
\hline
\end{tabular}%
\end{center}
\end{table}

For specification, in Fig. 2 we present $T$ (the modulus of
$\bm{T}$) as a function of $\theta$ from 0$\textordmasculine$
(zigzag) to 90$\textordmasculine$ (armchair), where the different
colored and shaped dots denote for the different combinations of
\emph{m} and \emph{n}. For example, the (red) spheres indicate the
value of $T$ for $|m-n|$=2, the (blue) squares for $|m-n|$=4, and so
on. Importantly, the (black) and (red) circles denote the values for
(2$m'$+1,1) and (1,2$n'$+1), which correspond to the directions with
chiral angle near 0 and 90$\textordmasculine$, respectively. As
shown in Fig. 2, the distribution of $T$ varying with $\theta$ for
the directions simultaneously obey $|m-n|$=2, (2$m'$+1,1) and
(1,2$n'$+1) forms an interesting pattern of two separated U-shapes.
All other points with different chiral indexes are located inside
the U-shapes, except for the three star-indicated special points.
One is the (green) star-indicated split-point between two U-shapes
at $\theta$=52.84$^{\circ}$ (SZ direction), along which
$T$=$(a_1^2+a_2^2)^{1/2}$. The other two (brown) and (blue) star
points at $\theta$=0$^{\circ}$ and 90$^{\circ}$ correspond to the
two boundaries of zigzag and armchair, respectively. Further, the
bottoms of two U-shapes are at the two special chiral angles of
23.74$^{\circ}$ and 75.82$^{\circ}$, respectively, which correspond
to the SA and newly discovered SL directional boundaries. Moreover,
with the increase of $|m-n|$, $T$ become longer and and the atomic
configuration of the corresponding boundary turn out to be more
complicated.

Based on the above description we conclude that, as shown in Fig. 1
and Tab. I, a phosphorene can essentially be divided into four major
regions as the chiral vector rotates around the $z$-axis from the
zigzag direction at chiral angle $\theta$=0$^{\circ}$ to the
armchair at 90$^{\circ}$. And these four regions are also indicated
by the angle ranges of 0-$\theta_1(23.74\textordmasculine)$,
$\theta_1$-$\theta_2(52.84\textordmasculine)$,
$\theta_2$-$\theta_3(75.82\textordmasculine)$ and
$\theta_3$-90$\textordmasculine$, with the regional boundaries
determined by the five crystallographic directions along zigzag, SA
[(blue) thick solid line], SZ [(green) thick solid], SL [(red) thick
solid] and armchair, respectively. More specifically, the
crystallographic directions of the reginal boundaries are described
by indexes (2,0), (3,1), (1,1), (1,3) and (0,2) respectively at
$\theta$=0, $\theta_1$, $\theta_2$, $\theta_3$ and 90$^{\circ}$, or
equivalently by the unitary vector length $T$=$a_1$,
$(9a_1^2+a_2^2)^{1/2}/2$, $(a_1^2+a_2^2)^{1/2}/2$,
$(a_1^2+9a_2^2)^{1/2}/2$ and $a_2$. We notice that, excepting for
the new SL, the other four peculiar directions and their
corresponding PNRs have been previously studied \cite{Marko, Ashwin,
Liu}. The PNRs cutting along SL direction also with stable edges,
however, has not been reported yet.

Further, there also exist other crystallographic directions within
each region, which are exemplarily shown in Fig. 1 by the thin solid
lines. Along these directions the projection of unitary vector
$\bm{T}$ on the horizontal (vertical) axis can be obtained by
combination of $a_1$/2 ($a_2$/2). For instance, when the chiral
angle rotates anticlockwise from $\theta_1$ with ($m,n$)=(3,1), the
vertical component of $\bm{T}$ is always kept as $a_2$/2 since the
data are in the left periphery of the first 'U' (see Fig. 2), and
the horizontal one can be added with integer times of $a_1$, such as
(5,1) shown by the (blue) thin solid lines shown in Fig. 1. And the
subsequent possible crystallographic directions in this region are
listed in the first column of Tab. I. Similarly, when the chiral
angle rotates anticlockwise from $\theta_1$ to $\theta_2$ (SA to
SZ), the corresponding horizontal (vertical) component of $\bm{T}$
varies in times of $a_1$/2 ($a_2$/2), such as (4,2). And a series of
possible directions are also presented in the second column of Tab.
I. However, according to the above derivation rule, we can not get
the varying law for boundaries from SZ to the armchair direction.
Therefore, to complete phase diagram, the new SL directional
boundary must be filled in the region from SZ to the armchair. When
the chiral angle rotates clockwise from $\theta_2$ to $\theta_3$ (SL
to SZ), the horizontal (vertical) projection of $\bm{T}$ varies with
times of $a_1$/2 ($a_2$/2). In contrast, as the SL direction rotates
anticlockwise to the armchair there is a change of multiple $a_2$
for the vertical component of $\bm{T}$ but the horizontal one
maintains unchanged due to the data for the boundaries are
distributed in the right periphery of the second U. The resulting
series of boundary chiral numbers are shown in the third and fourth
columns in Tab. I, respectively. The examples of the crystal
directions with chiral indexes (4,2), (2,4) and (1,5) from
$\theta_1$ to 90$\textordmasculine$ are indicted in Fig. 1 by the
(blue) and (red) thin solid lines, respectively. However, as shown
by the colored stars in Fig. 2, the data for the three special
boundary directions, zigzag, SZ and armchair, are outside the
U-shape valleys. In fact the nanostructures with these edges are
most stable and favorably formed in experiments \cite{Liang}.

Moreover, when we cut a 2D phosphorene sheet along a certain crystal
direction, in fact, there exist two different manners for tangent to
atoms due to the two nonequivalent phosphorus atoms per primitive
cell. Therefore, the direction (1,1)52.84$\textordmasculine$ is very
particular, behind/beyond along which the boundary edge is
non-beard/beard when tangent to the lower atom, and viceversa when
tangent to the upper atom of the primitive cell. That is to say, SZ
is the demarcation line between with and without beard for the
boundary edges. Further, for cutting manner by the solid lines in
Fig. 1, all the boundary edges have an obvious zigzag-like edge
atomic morphology except along the SZ and armchair directions.
Whether a boundary edge is bearded or non-beard depend on the number
of dangled bonds of the outermost atoms. For beard/non-beard
boundary edge, the number of dangled bonds of the outermost atoms is
two/one per atom.

Actually, when we consider the phosphorene nanostructures the effect
of the $z$-direction on the characterization must be taken into
account. This effect can be reflected by the fact that there are two
kinds of arrangement for the outermost atoms along each
crystallographic direction. According to our definition for
$\bm{T}$, the chiral indexes satisfying \emph{m}+\emph{n}=even have
defined the all directions for a phosphorene sheet shown in Fig. 1.
When both \emph{m} and \emph{n} are odd, however, the outermost
phosphorus atoms along the direction are alternately located at the
upper and lower sublayers. So the supercell width of the nanoribbon
cutting along this direction is twice as $T$. The typical example of
this case is shown by the (red) thick solid line segment along
SA(3,1) direction in Fig. 1. In comparison, when both \emph{m} and
\emph{n} are even, the outermost atoms are located at the same
sublayer. In this case the supercell width is just $T$ [see the
(4,2) direction]. The detailed classification on nanoribbons is
discussed in the Supplemental Material \cite{SM}.

\subsection{PNRs and the PNR-based z-shaped junction}
According to the structural characterization on phosphorene in the
previous subsection, one can principally obtain PNRs with different
edge morphologies by cutting a phosphorene sheet along different
crystallographic directions. In addition to the extensively studied
ZPNR and APNR \cite{Carvalho, Ezawa}, the ribbons by cutting along
other three regional boundary directions SA(3,1), SZ(1,1) and
SL(1,3), indicated by the thick solid lines in Fig. 1, are
respectively denoted as (3,1)nbPNR, (1,1)bPNR and (1,3)bPNR, where
bPNR or nbPNR means a ribbon with or without beard edges. On the
other hand, when the ribbons by cutting along SA(3,1), SZ(1,1) and
SL(1,3) according to the dashed lines in Fig. 1 are denoted as
(3,1)bPNR, (1,1)nbPNR and (1,3)nbPNR, respectively. These six types
of PNRs are respectively sketched in Figs. 3(a-f), where the (red)
dashed parallelogram in each PNR indicates its minimum periodical
supercell. Since their outermost atoms along these ribbon edges are
alternately located at the upper and lower sublayers, their
supercell width is twice as $T$ for both beard and nonbeard edges.
Among these ribbons only (3,1)PNR and (1,1)PNR have been previously
studied \cite{Marko}. For (1,3)PNRs along SL, the chiral angle is
75.82$^{\circ}$ and the edge morphology is more likely to APNR,
which is obviously different from (3,1)PNRs along SA
(23.74$^{\circ}$) near zigzag edge. In addition, these ribbons
obviously do not have the mirror symmetry as the normal ZPNR and
APNR \cite{Carvalho}. As mentioned in the above subsection, the
ribbons cutting along SA and SL solid line have a periodic
zigzag-like edge morphology. Importantly, this edge morphology are
destroyed for the ribbons along the dashed line directions.
Therefore, for reflecting the particular edge atomic morphology, as
show in Figs. 3(a-f), we have to choose different (red) dash-line
parallelograms as the supercells for (3,1)nbPNR, (1,1)bPNR,
(1,3)bPNR, (3,1)bPNR, (1,1)nbPNR and (1,3)nbPNR, respectively.

\begin{figure}
\centering
\includegraphics[bb=30 51 503 348,width=8.5cm]{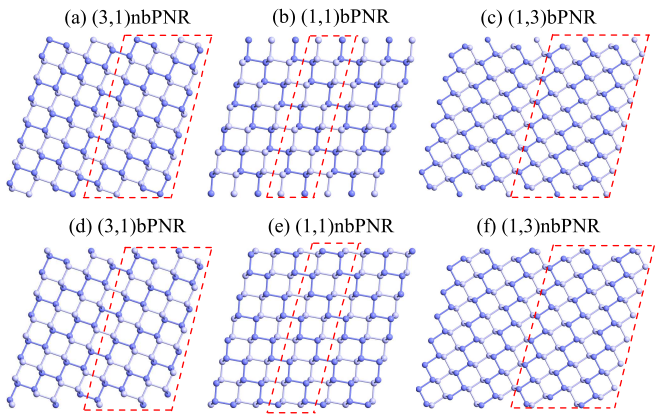}
\caption{The illustrations of (a) (3,1)nbPNR, (b) (1,1)bPNR, (c)
(1,3)bPNR, (d) (3,1)bPNR, (e) (1,1)nbPNR and (f) (1,3)nbPNR, where
the (red) dashed parallelogram in each PNR indicates its minimum
periodical supercell.}
\end{figure}

The stability of PNRs is an important issue which has been
previously studied by calculating the free energy \cite{Han,
Ashwin}. Since the stability for (3,1)PNR and (1,1)PNR has been
studied in previous work which implies that the non-beard ribbon is
more stable. So in this work, we just compare the stability of the
three ribbons in the case of non-beard, and also study the
difference in stability between (1,3)bPNR and (1,3)nbPNR. Using the
first-principles calculation based on the density functional theory
(DFT) \cite{Ren}, we have verified the binding energies of the four
narrow ribbons shown in Figs. 3(a, c, e, f). The order of the
stability for these ribbons is (1,1)nbPNR, (3,1)nbPNR, (1,1)nbPNR,
(1,3)nbPNR, (1,3)bPNR, with width $\sim$2 nm. With the increase of
width, the ribbons will become more stable with lower binding
energies.

\begin{figure}[t]
\centering
\includegraphics[bb=6 52 607 342, width=8.5cm]{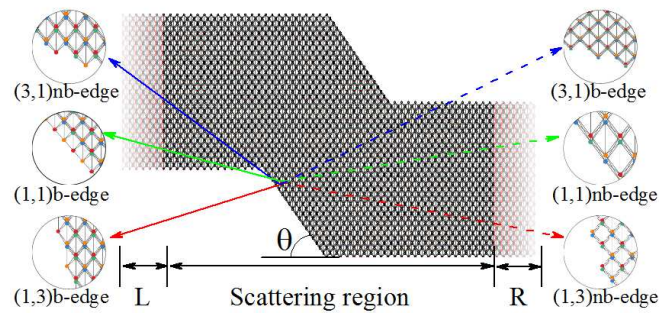}
\caption{Schematic illustration of the PNR-based z-shaped junction,
where the left and right enlarged view for the six types of edge
geometric morphologies in the central scattering region as the
variation of the tilting angle [corresponding to Figs. 3(a-f)].}
\end{figure}

Next, we propose a two-terminal z-shaped junction based on PNRs as
shown in Fig. 4, where the electrodes of the junction are two
semi-infinitive ZPNRs with the same width $\sim$100 nm and the
central scattering region is a z-shaped phosphorene quantum dot
(PQD) which contain a segment of PNR with different tilting angles.
The tilting angle $\theta$ of PQD corresponding to the chiral angle
varies from 0 to 90$^{\circ}$, as a result the two hypotenuse edges
of PQD sweep over all possible geometrical morphologies. The
enlarged typical edge geometry of the ribbons in Figs. 3(a-c)/(d-f)
are shown in the left/right side of Fig. 4. This type of homogeneous
junctions may be etched on a sample sheet of 2D materials as the
development of experimental technology, and it has been extensively
studied for graphene, silicene, $h$-BN, and so on \cite{Lee,
WangZF}. However, for phosphorene the edge (state) effect has only
been considered for quantum dot and ring structures \cite{ZhangR,
ZhangR1}, although the PNR-based z-shaped junction has been
primarily studied \cite{Enrique}. In this work, we further study
this z-shaped PNR junction to explore the influence of the variation
in edge geometrical morphology (corresponding to different tilting
angle) on the transport property. The significance is that the
connection between the different parts in homogeneous junction does
not destroy the original lattice structure. This may be impossible
for inhomogeneous heterojnuctions with another material electrodes
because there is always a lattice mismatch. Furthermore, the
z-shaped junction allows short ribbon with all possible edges
embedded in the junction, in which the existence of resonant
tunneling phenomenon \cite{Katsunori} can be expected.

\subsection{Computational details}
The numerical calculations here are performed by using Kwant code
based on the tight-binding (TB) Hamiltonian, and the atomistic
quantum transport simulations are based on the scattering-matrix
approach from solving wavefunction
\cite{Yongjin,Zwierzycki,Tatiane}. Compared with the conventional
first-principles calculations \cite{Ren}, this method can calculate
large nanostructures matching the usual experimental reachable
sample size up to sub-100 nm scales with better precision
\cite{Groth}. This advantage enables us to truly explore the
electron structure and transport property for the phosphorene
nanostructures with different edge morphologies realized in
experiments, and to provide a fundamental understanding in
preparation for the design of nano-devices \cite{Lill,Anffany}.

In specification, the ribbons shown in Fig. 3 and the junction shown
in Fig. 4, in which the system with and without beard edges is
respectively considered, and the stability is analyzed by the
first-principles approach. In the calculation the scale of the
systems is adapted about 10-30 nm to meet the current experimental
development \cite{Anffany}, which is large enough to show a good
illustration of the credibility and feasibility for the calculated
results. This scale is more suitable for using Kwant, a Python
package for numerical quantum transport calculations. We use a TB
model based on the quasi-particle orbit to describe the low-energy
electrons in phosphorene \cite{Rudenko}. The parameters of the TB
model are obtained by fitting its low-energy band structure of
phosphorene to the one computed by the DFT-GW approach
\cite{Rudenko}, and it was shown that the fitted TB model can
reproduce well the band structure of phosphorene in the low-energy
regime as compared to the DFT-GW approach \cite{Rudenko}.

The formal TB Hamiltonian for the considered system is
\begin{equation}
H=\sum\limits_{i}\varepsilon_{i}c_{i}^{\dag}c_{i}+\sum\limits_{i{\neq}j}t_{ij}c_{i}^{\dag
}c_{j},
\end{equation}
where the summation runs over all lattice sites, $\varepsilon_{i}$
is the on-site energy at site $i$, $t_{ij}$ is the hopping energy
between sites $i$ and $j$, and $c_{i}^{\dag}$ ($c_{j}$) is the
creation (annihilation) operator of an electron at site $i$ ($j$).
It has been shown that five hopping parameters (see Fig. 1) are
enough to describe the electronic band structure of phosphorene with
hopping energies $t_1$=-1.220 eV, $t_{2}$=3.665 eV, $t_{3}$=-0.205
eV, $t_{4}$=-0.105 eV, and $t_{5}$=-0.055 eV \cite{Rudenko}.
Starting from the Hamiltonian, we can calculate the local density of
states (LDOS) using the following formula
\begin{equation}
\text{LDOS}(E)=\frac{1}{c\sqrt{2\pi}}\sum\limits_{n}|\Psi_{n}(r)|^{2}e^{\frac{-(E_{n}-E)^{2}}{2c^{2}}},
\end{equation}
where $c$ is broadening parameter, $\Psi_{n}(r)$ and $E_{n}$ are the
eigenfunction and eigenvalue, respectively, in which $n$ denotes the
energy band index and $r$ the atom position. Further, the
conductance of the junction at zero temperature can be calculated by
using the Landauer formula \cite{Datta}
\begin{equation}
G=\frac{2e^{2}}{h}\int{dE(-\frac{\partial{f_{0}}}{\partial{E}})T(E)},
\end{equation}
where $f_0(E)$=$1/{[e^{{(E-E_F)}/{k_{B}T}}+1]}$ is the equilibrium
distribution function with Fermi energy $E_F$, L/R labels the
left/right electrode of the system, and $T(E)$=
$\text{Tr}[\Gamma_L(E,V)G (E,V)\Gamma_R(E,V)G^{\dag}(E,V)]$ is the
transmission coefficient through the junction from electrode $L$ to
$R$, in which $G(G^{\dag})$ is the retarded (advanced) Green
function of the central region, and $\Gamma_{L/R}$ the coupling
matrix between the scattering region and the left/right electrode.

\section{Numerical Results and Discussions}

\subsection{Electronic structure for PNRs}
By solving the difference Schr\"{o}dinger equation corresponding to
Hamiltonian on the proper basis for the supercell adopted in Fig. 3
and applying the Bloch theorem, the $k$-dependent Hamiltonian for a
PNR can be written as
$H(k)$=$H_{0,0}+H_{0,1}e^{ika}+H_{0,1}^{\dag}e^{-ika}$ in the form
of ($N\times N$) dimensional matrix. Here $N$ is the number of atoms
in the supercell drawn by the (red) dashed line in Fig. 3, $H_{0,0}$
is the matrix of the central cell, $H_{0,1}$ the coupling matrix
with the right-hand adjacent cell, and $a$ is the length between two
nearest-neighbor cells. Diagonalizing this $k$-dependent
Hamiltonian, we can obtain the band spectrum and the corresponding
eigen-wavefunctions. And following the procedure described in the
last subsection, we can also obtain the conductance and LDOS for the
system.

\begin{figure}[b]
\centering
\includegraphics[bb=5 1 539 436,width=8.5cm]{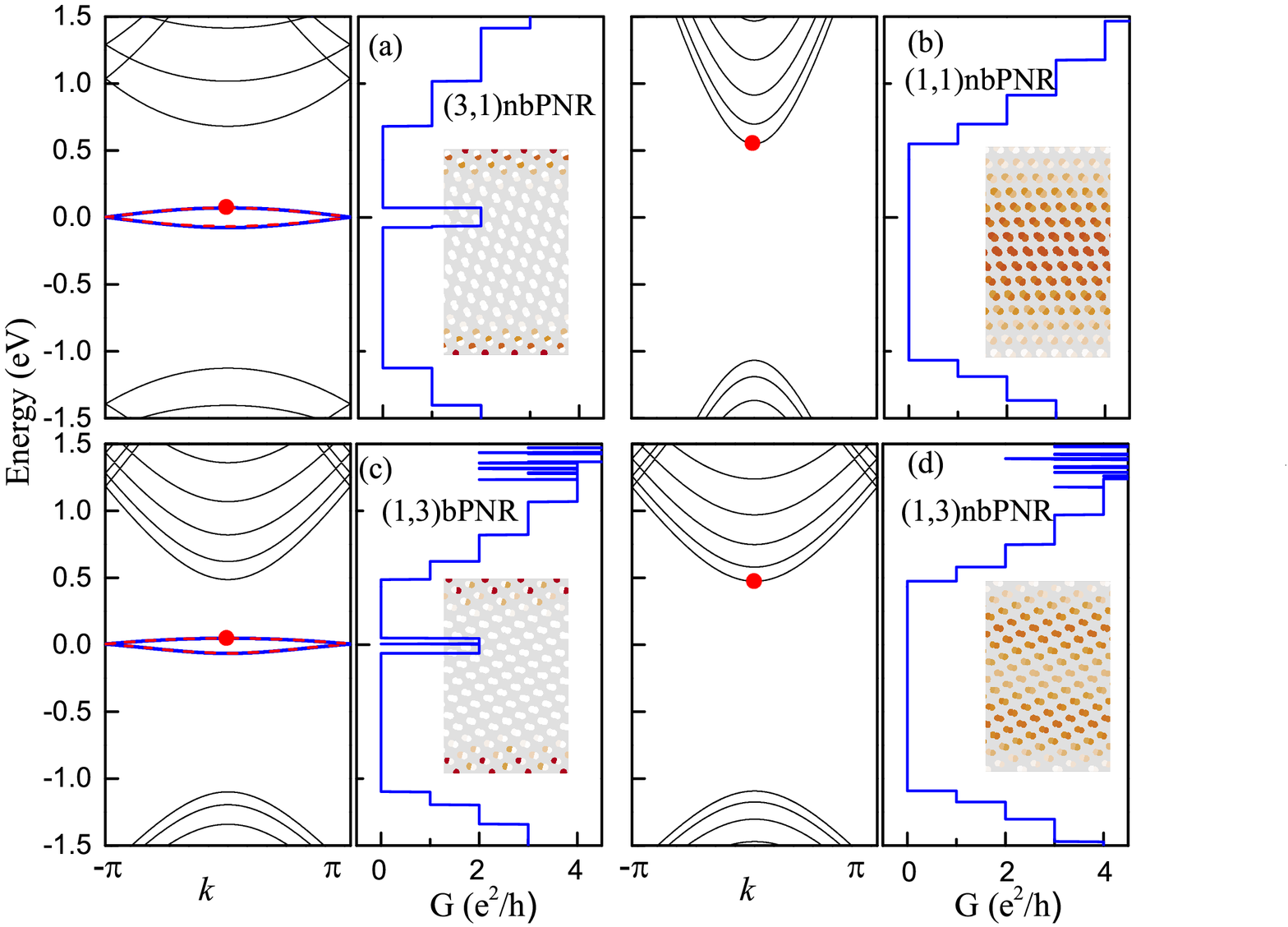}
\caption{The calculated energy band (left panel) and conductance
(right panel) for (a) (3,1)nbPNR, (b) (1,1)nbPNR, (c) (1,3)bPNR and
(d) (1,3)nbPNR with width around 4 nm, respectively, where the
embedded inset in each right panel is the LDOS at energy indicated
by the (red) dot in the corresponding left panel.}
\end{figure}

The calculated energy band (the left panel) and conductance (right
panel) are shown in Figs. 5(a-d) for the four most stable
(3,1)nbPNR, (1,1)nbPNR, (1,3)bPNR and (1,3)nbPNR with width around 4
nm, respectively. And the choice of the supercells has consequently
taken the effect of the $z$-direction into consideration. In each
figure the embedded inset in the right panel is the LDOS at energy
indicated by the (red) dot in the corresponding left panel. First,
for the band structure of (3,1)nbPNR shown in Fig. 5(a), the
twofold-degenerate edge bands indicated by the (red) dashed and
(blue) solid lines pass through the Fermi level exhibiting metallic
property, while (1,1)nbPNR is a direct band gap semiconductor as
shown in Fig. 5(b). This recovers the results by previous
straightforward calculation for these two types of ribbons
\cite{Marko}. However, for the newly discovered (1,3)bPNR the
twofold-degenerate edge bands around the Fermi level are also
detached from the bulk bands. This is similar to the band structure
of (3,1)nbPNR, but the bulk band (black thin solid line) and its
position relating to edge state are still significantly different.
To further confirm the existence of the edge states, we have
calculated the LDOS at the energy marked by the (red) dots on edge
bands, the thumbnail of LDOS embedding in the right panels depicts
the real-space localization of electrons. From the LDOS for
(3,1)nbPNR and (1,3)bPNR, we observe the obvious electron clouds
mainly localized at the unsaturated phosphorus atoms in the
zigzag-like edges of the ribbons, which results from the lone pair
electrons at the edges \cite{Tran}. In contrast, for (1,1)nbPNR the
electron states at the conduct band minimum (CBM) is not contributed
by the atoms at the edge of ribbons.

Further, the distinctive electronic properties of PNRs have a direct
consequence on the transport. As shown in each right panel of Figs.
5(a-d), the conductance displays a clear stepwise structure. The
quantized conductance plateaus follow the sequence $G=ne^{2}/h$,
with $n$ denoting positive integers. For (3,1)nbPNR and (1,3)bPNR,
when energy is within the band gap around the Fermi level, their
conductance have a quantized plateau, which is clearly induced by
edge states. However, both (1,1)nbPNR and (1,3)nbPNR do not have a
quantized plateau near the Fermi level because there are no edge
states near the Fermi level. On the other hand, we need to
understand the effect of a PNR with beard or non-beard edge on the
electronic structure. Previous studies have shown that for the
beard-ZPNR [or (2,0)bPNR], (3,1)bPNR and (1,1)bPNR, the band
structure is essentially different from that for the non-beard
counterpart, especially whether or not existing edge states
\cite{Ezawa, Marko}. For (2,0)bPNR and (3,1)bPNR, we see that the
bulk bands remain unchanged, but the edge states are disappeared.
For the new (1,3)bPNR, an edge band occurs around the Fermi level.
However, (1,3)bPNR is special due to the original beard edges, and
its band structure is characterized by the significant edge states,
as shown in Figs. 3(c) and 5(c). Interestingly, when we translate
the edge outwards by one atom for (1,3)bPNR, we then obtain
(1,3)nbPNR, as shown in Fig. 3(f). As a result, as shown in Fig.
5(d) in comparison with 5(c), although its bulk bands are unchanged
the edge states are disappear. Hence it becomes a direct band gap
semiconductor like APNR and (1,1)nbPNR.

For a pristine PNR, in summary, with or without beard edge
determines the existence of the zigzag-like edge morphology
resulting in the edge state bands. The electronic localization
caused by the edge zigzag-like atomic configuration may be from that
phosphorene is a second-order topological insulator \cite{Ezawa2},
which is still an open question. The PNRs with possible edges along
the directions by solid lines in Fig. 1 all have edge states near
the Fermi energy except for armchair-edge. Since along these
directions there is always a periodic zigzag-like edge geometry,
except for (1,1)bPNR whose edge states are caused by the unsaturated
atoms. Therefore, the (1,1)SZ crystallographic direction is a
demarcation line, behind/beyond along which the PNRs without/with
beard edges, or viceversa for an alternative tangent manner along
the same direction. Within the chiral angle range of of 0-$\theta_2$
(below SZ), all the PNRs without beard edge have a zigzag-like edge
morphology, but this edge morphology is absent for the corresponding
beard ones. In this case the electrons are not localized on the
outmost atoms, which result in no edge states. For the
$\theta_2$-90$\textordmasculine$ range, in contrast, the zigzag-like
edge morphology of beard PNRs is originally from the dangling
phosphorus atoms, but this property for non-beard ones is
disappeared due to the saturation of the dangling atoms. This
conclusion is consistent with that for graphene \cite{Akhmerov}.

\subsection{Transport property of the z-shaped junction}
\begin{figure}[b]
\centering
\includegraphics[bb=1 2 614 410, width=8.6cm]{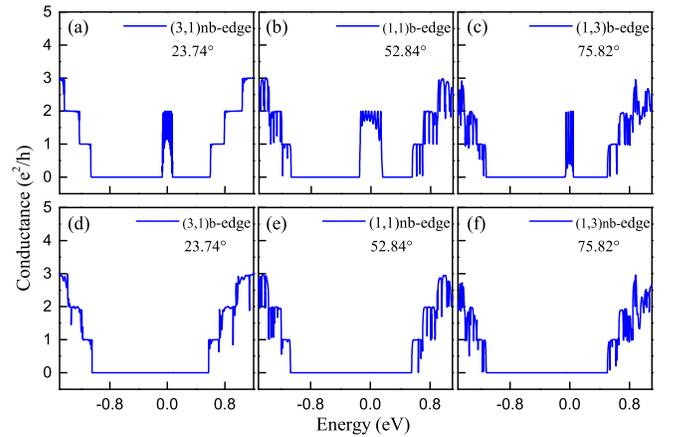}
\caption{The conductance (in units of $e^2/h$) spectrum of the
z-shaped junction, where the central scatter PQD with (a)
(3,1)nb-edge, (b) (1,1)b-edge, (c) (1,3)b-edge, (d) (3,1)b-edge, (e)
(1,1)nb-edge and (f) (1,3)nb-edge corresponding to Figs. 3(a-f),
respectively.}
\end{figure}

\begin{figure*}
\centering
\includegraphics[bb=11 11 1007 498,width=17cm]{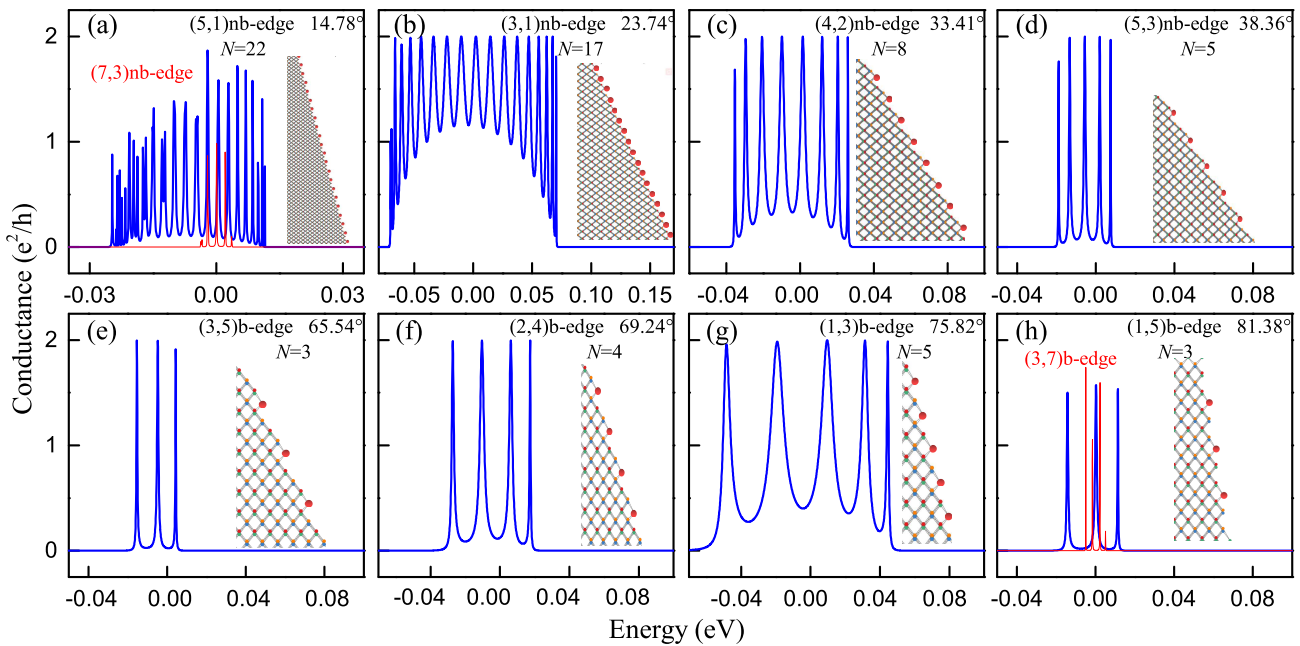}
\caption{The oscillation peaks on conductance terrace (in units of
$e^2/h$) within low energies for the z-shaped junction with the
increase of titling (chiral) angle $\theta$, where (a) for
(5,1)nb-edge, (b) (3,1)nb-edge, (c) (4,2)nb-edge, (d) (5,3)nb-edge,
(e) (3,5)b-edge, (f) (2,4)b-edge, (g) (1,3)b-edge and (h)
(1,5)b-edge, and the (red) thin line in (a)/(h) is the conductance
for the (7,3)nb-/(3,7)b-edge. The insets show the corresponding PQD
edge geometries, where the (red) dots denote the outmost phosphorus
atoms along the zigzag-like edge, and integer $N$ is the number of
the outmost atoms.}
\end{figure*}

The proposed z-shaped junction in Fig. 4 consists two semi-finite
ZPNR electrodes and the central scattering region of a z-shaped PQD.
As the degree of torsion of junction is varied, the tilting angle of
the center PQD can be altered from 0 to 90$^{\circ}$, hence the
whole junction is changed from a standard ZPNR to APNR. However, in
this process the two edges of PQD sweep over all possible
geometrical morphologies described in Sec. IIB (also see Fig. 1 and
Tab. I). As a result the electronic band structure of the PQD varies
alternatively between with and without edge states near Fermi
energy, and the whole junction is consequently either conducting or
semiconducting. In order to explore the variation of the transport
for the junction, in Figs. 6(a-f) we first show the calculated
conductances (in units of $e^2/h$) for the junction with PQD
edges along the three regional boundaries shown in the two sides of
Fig. 4, with (3,1)nb-, (1,1)b-, (1,3)b-, (3,1)b-, (1,1)nb- and
(1,3)nb-edge, respectively. Apparently, as shown in Figs. 6(a-c),
the junction of PQD with (3,1)nb/(1,1)b/(1,3)b-edge exhibits a
conductance plateaus of $2e^2/h$ with oscillations within the energy
range -0.072/-0.153/-0.095 to 0.072/0.156/0.053 eV due to the edge
states \cite{Marko}. On the contrary, the junction of PQD with
(3,1)b-, (1,1)nb- and (1,3)nb-edge, as shown in Figs. 6(d-f),
displays almost zero conductance because of the obvious band gap
\cite{Marko}. The result indicates that the edge states of the PQD
also play an important role in transport for a z-shaped junction.
This interesting transport property has also been reported for a
planar homogenous junction formed by graphite ribbons with different
edges \cite{Katsunori}.

Next, to find the origin of the conductance oscillating peaks, in
Fig. 7 we present the low-energy conductance spectra for the
junction with $\theta$ in ascending order from 14.78 to
81.38$\textordmasculine$, within which the PQD along directions of
(5,1)nb-, (3,1)nb-, (4,2)nb-, (5,3)nb-, (3,5)b-, (2,4)b-, (1,3)b-
and (1,5)b-edge are selected, respectively. These eight edges all
have an obvious zigzag-like geometry with the enlarged view of edges
inserted in each figure. Consistent with the previous the band
structure for PNRs with different edges, in Figs. 7(a-d) the
conductance spectra for the PQD with four different nonbeard edges
all have an oscillating terrace provided by the edge states around
the Fermi level. Since the function of the outermost atoms along the
zigzag-like configuration are similar to those of actual
zigzag-edge, so they are used as a medium atom for electron
transport \cite{Ezawa}. On the other hand, the spectra shown in
7(e-h) respectively for beard (3,5), (2,4), (1,3) and (1,5) edges
have a similar feature with an oscillating terrace provided by the
edge states. Although these edges with tilting angle close to
90$\textordmasculine$ resulting in the edge atomic configuration
similar to armchair type, for the beard SL and its derivative edges
still form a periodic zigzag-like configuration due to the outermost
dangled phosphorus atoms. This function of the beard edge for PNRs
has also been demonstrated previously \cite{Ezawa, Marko}.

When the hypotenuse of PQD with edge behind/beyond SZ direction,
such as (3,1)nb- and (1,3)b-edge, the junctions exhibit similar
transport behavior due to existence of zigzag-like edge geometry.
This is because that the outermost phosphorus atom along the
zigzag-like edges of PQD form an electronic bound-state
\cite{Ezawa2}, which is expressed in the energy band as discrete
energy levels. These discrete energy levels cause the orphaned
electrons at the edge to be hopped to produce an oscillating
transport behavior. The function of the outermost atoms along edges
can also be well reflected in the conductance spectrum.
Interestingly, as clearly shown in Figs. 7(a-h), the number ($N$) of
the terrace oscillating peaks exactly matches the number of the
outermost phosphorus atoms indicated by the (red) dots within an
hypotenuse edge of PQD, which is also related to the number of
minimum period length $T$ along a hypotenuse edge of PQD because
that the electrons are localized at each zigzag configuration
regardless of which sublayer. Therefore, for a z-shaped junction
with certain size, as the edge of the central PQD continuously
sweeps over all possible $\theta$ the value of $N$ is varied. In
addition to the above eight tilting angles, there are many other
angles as indicated by the different colored and shaped dots in Fig.
2. The conductance spectrum of the junction with these edge
boundaries also exhibits oscillating characteristics in the low
energy regime, as shown by the (red) thin lines in Figs. 7(a) and
7(h). For large $T$, the zigzag-like edge may have more inner
structures in comparison with those with edges for small $T$.
Further, these atomic configurations also produce various scatters
that can cause severe interference to the oscillating transport of
the electrons, making the discrete levels degenerate and reducing
the corresponding conductance. For example, in Fig. 7(a), due to the
degeneracy of the energy level and the scattering of the beard edge
atomic configurations, the number of conductance peaks is less than
that of the outermost phosphorus atoms, and the conductance value is
also less than $2e^{2}/h$. On the other hand, the value of $T$ also
limits the conductance. By more calculations, we find that the
conductance of the junctions with $T$ less than 20 $\AA$ are
effective in the low energy regime, otherwise the conductance value
is extremely small as $T$ is beyond this critical value.

Finally, the data of the calculated conductance vs titling angle for
our z-shaped junction whose central PQD with more possible edges
with $T$ less than critical value 20 $\AA$ are summarized in Fig. 8,
where the blue (red) triangles (spheres) for the non-beard (beard)
case and the vertical dashed lines indicate the three particular
regional boundaries. As the titling (chiral) angle $\theta$ varying
from 0 to 90$\textordmasculine$, we can see that the SZ(1,1) at
52.84$\textordmasculine$ is more special, behind/beyond which the
conductance is nonzero/zero for the junction with non-beard edges.
The situation for the junction with beard edges is just in opposite
to the non-beard case, and the two cases are complementary. In
addition, more information can be obtained from the phase-like
diagram shown in Fig. 8. For instance, the conductance of junction
with $\theta$ range from 0 to 90$\textordmasculine$ is between 0 and
2$e^2$/h sensitively depending on the module length $T$. Since near
the three regional boundaries 0$\textordmasculine$ (zigzag),
52.84$\textordmasculine$ (SZ) and 90$\textordmasculine$ (armchair)
$T$ is more larger, which results in more smaller conductance. This
result is in consistence with the distribution of $T$ near the
bottom of U in Fig. 2. If $T$ is beyond this critical value, the
conductance tends to zero although there are still edge channels for
the junction in the low energy regime.

\begin{figure}
\centering
\includegraphics[bb=8 0 611 519, width=8.5cm]{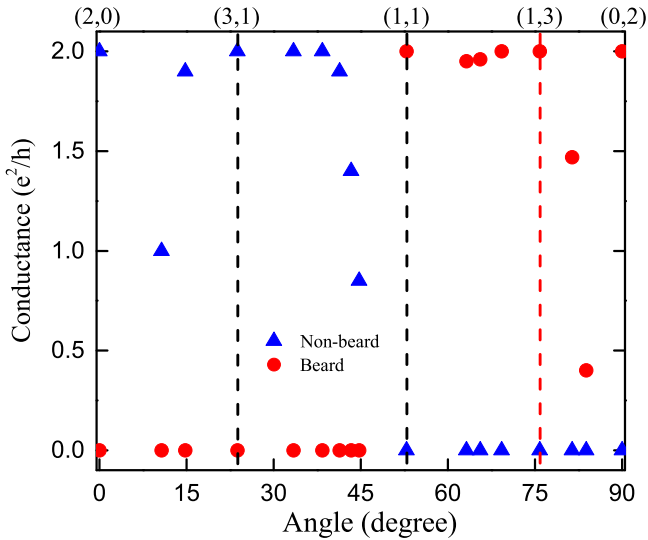}
\caption{The conductance of the junction as a function of tilting
angle with $|\bf{T}|$ less than 20 \AA, where the hypotenuse of the
central part with and without beard edge are indicated by the (red)
spheres and the (blue) triangles, respectively, and the vertical
dashed lines imply the three particular regional boundaries.}
\end{figure}

\section{Summary and Conclusion}
We have theoretically studied  the 2D crystallographic
characterization on an anisotropic phosphorene and its applications
in nanostructures. A chiral number (vector) or chiral angle is
defined to uniformly describe the planar crystallographic directions
on a phosphorene. As the chiral angle increasing from
0$\textordmasculine$ to 90$\textordmasculine$, the along boundary
edge geometry varies from zigzag to armchair and sweep over all
possible edge morphologies during the process. We find a phosphorene
can be divided into four fan-shaped regions determined by boundary
chiral angles 0$\textordmasculine$,
$\theta_1(23.74\textordmasculine)$,
$\theta_2(52.84\textordmasculine)$,
$\theta_3(75.82\textordmasculine)$ and 90$\textordmasculine$, among
which the characteristics is different each other. Therefore, by
cutting a phosphorene along these five regional boundary directions,
one can obtain previously studied ZPNR, SAPNR, SZPNR, and APNR.
Importantly, the new type of PNR with SL edge geometry along
$\theta_3$ has been discovered. Essentially, the SL direction is
similar to SA one with a periodic zigzag-like edge morphology
supporting edge states. Further, we then demonstrate the variation
in electronic property for PNRs with different edge atomic
geometries by using the Kwant tight-binding (TB) code, in which a
particular attention is paid to the new SLPNR intrinsically with
twofold-degenerate edge states. In addition, in each regional
boundary direction the origin of beard and non-beard edges is
discussed and related to the existence of edge states. We find that
the (1,1) direction is the dividing line, below/beyand which the
ribbon with nonbeard/beard edges.

Second, we propose a two-terminal z-shaped homogenous junction based
on PNR and calculate its transport property by using the scattering-matrix for treating large enough systems comparable with
experimental samples. The two electrodes of the junction are
semi-infinitive ZPNRs with the same width $\sim$100 nm, where the
central scattering region of a segment of PNR with edge tilting
angles from 0 to 90$\textordmasculine$ is sandwiched between them.
In consequence the two hypotenuse edges of the central PQD sweep
over all possible geometrical morphologies. Interestingly, the
conductance spectrum sensitively depends on the tilting angle of
z-shaped junction. For the junctions of PQD with edge states the
transport exhibit a conductance terrace with resonant oscillation in
the low energy regime, otherwise the electron transport is blocked
due to the absence of edge states. Remarkably, the number of
conductance oscillating peaks is exactly equal to the number of
outermost phosphorus atoms in the zigzag-like edges of the center
PQD, since the discrete energy levels of the zigzag-like atomic
configurations provide the transport channels for electrons. The
findings here may be extended to the group-VA 2D materials, such as
2D group-VA arsenene, antimonene and bismuthene, for observing the
electronic property of nanostructures in experiments, and provide a
reference for the preparation of better nano-devices.

\begin{acknowledgments}
This work was supported by the National Natural Science Foundation
of China (Grant Nos. 11774085, 11804092, 11704118), China
Postdoctoral Science Foundation funded project (Grant Nos.
BX20180097, 2019M652777), and Hunan Provincial Natural Science
Foundation of China (Grant No. 2019JJ40187).
\end{acknowledgments}

\end{document}